\documentclass[12pt]{article}
\usepackage{axodraw,bbold}

\catcode`@=12
\begin{document}
\thispagestyle{empty}
\begin{flushright} 
UCRHEP-T395\\ 
August 2005\
\end{flushright}
\vspace{0.5in}
\begin{center}
{\LARGE	\bf Tetrahedral Family Symmetry\\
and the Neutrino Mixing Matrix\\}
\vspace{1.5in}
{\bf Ernest Ma\\}
\vspace{0.2in}
{\sl Physics Department, University of California, Riverside, 
California 92521, USA,\\
and Institute for Particle Physics Phenomenology, Department of Physics,\\ 
University of Durham, Durham, DH1 3LE, UK\\}
\vspace{1.5in}
\end{center}

\begin{abstract}\
In a new application of the discrete non-Abelian symmetry $A_4$ using 
the canonical seesaw mechanism, a three-parameter form of the neutrino 
mass matrix is derived.  It predicts the following mixing angles for 
neutrino oscillations: $\theta_{13}=0$, $\sin^2 \theta_{23}=1/2$, and 
$\sin^2 \theta_{12}$ close, but not exactly equal to 1/3, in one natural 
symmetry limit.
\end{abstract}

\newpage
\baselineskip 24pt

The symmetry group of the tetrahedron is also that of the even permutation 
of four objects, i.e. $A_4$.  It is a non-Abelian finite subgroup of 
$SO(3)$ as well as $SU(3)$.  It has twelve elements and four irreducible 
representations: \underline {1}, \underline {1}$'$, \underline {1}$''$, and 
\underline {3}.  It has been shown to be useful in describing 
\cite{mr01,bmv03,m0203,m04,m05,af05,cfm05,bh05} the family structure of 
quarks and leptons.  In most previous applications, the lepton doublets 
$(\nu_i,l_i)$ are assigned to the \underline{3} representation of $A_4$, 
whereas the charged-lepton singlets $l^c_i$ are assigned to the three 
inequivalent one-dimensional representations \underline{1}, \underline {1}$'$, 
\underline {1}$''$.  Here as in the two papers of Ref.~\cite{cfm05}, both 
$(\nu_i,l_i)$ and $l^c_i$ are \underline{3} instead.

Three heavy neutral fermion singelts $N_i$ are assumed, transforming as 
\underline{1}, \underline {1}$'$, \underline {1}$''$ under $A_4$. [In the 
first paper of Ref.~\cite{cfm05}, they transform as \underline{3}; in the 
second, they are absent.] The multiplication rule $\underline{1}' \times 
\underline{1}'' = \underline {1}$ implies that the Majorana mass matrix of 
$N_i$ invariant under $A_4$ is given by
\begin{equation}
{\cal M}_N = \pmatrix{A & 0 & 0 \cr 0 & 0 & B \cr 0 & B & 0}.
\end{equation}
The multiplication rule
\begin{equation}
\underline{3} \times \underline{3} = \underline{1} + \underline{1}' + 
\underline{1}'' + \underline{3} + \underline{3}
\end{equation}
allows the charged-lepton mass matrix to be diagonal by having three 
Higgs doublets transforming as \underline{1}, \underline{1}$'$, 
\underline{1}$''$, resulting in a diagonal ${\cal M}_l$ with
\begin{equation}
\pmatrix{m_e \cr m_\mu \cr m_\tau} = \pmatrix{1 & 1 & 1 \cr 
1 & \omega & \omega^2 \cr 1 & \omega^2 & \omega} \pmatrix{h_1 v_1 \cr h_2 v_2 
\cr h_3 v_3},
\end{equation}
where $\omega = \exp (2\pi i/3)$ and $v_{1,2,3}$ are the vacuum expectation 
values of these three Higgs doublets.

As for the Dirac mass matrix linking $\nu_i$ to $N_j$, three other Higgs 
doublets are assumed, transforming as \underline{3} under $A_4$.  [They 
are distinguished from the previous three Higgs doublets by a discrete 
$Z_2$ symmetry.] Thus
\begin{eqnarray}
{\cal M}_D = \pmatrix{f_1 u_1 & f_2 u_1 & f_3 u_1 \cr f_1 u_2 & f_2 \omega 
u_2 & f_3 \omega^2 u_2 \cr f_1 u_3 & f_2 \omega^2 u_3 & f_3 \omega u_3} 
= \pmatrix{u_1 & 0 & 0 \cr 0 & u_2 & 0 \cr 0 & 0 & u_3} 
\pmatrix{1 & 1 & 1 \cr 1 & \omega & \omega^2 \cr 1 & \omega^2 & \omega} 
\pmatrix{f_1 & 0 & 0 \cr 0 & f_2 & 0 \cr 0 & 0 & f_3}.
\end{eqnarray}
Using the canonical seesaw mechanism \cite{seesaw}, the Majorana neutrino 
mass matrix is then given by
\begin{eqnarray}
{\cal M}_\nu = {\cal M}_D {\cal M}_N^{-1} {\cal M}_D^T 
= \pmatrix{u_1 & 0 & 0 \cr 0 & u_2 & 0 \cr 0 & 0 & u_3} \pmatrix{a & b & b 
\cr b & a & b \cr b & b & a} \pmatrix{u_1 & 0 & 0 \cr 0 & u_2 & 0 \cr 0 & 0 
& u_3},
\end{eqnarray}
where
\begin{equation}
a = f_1^2/A + 2 f_2 f_3/B, ~~~ b = f_1^2/A - f_2 f_3/B,
\end{equation}
and $u_{1,2,3}$ are the vacuum expectation values of the second set of Higgs 
doublets which transform as \underline{3} under $A_4$.

If $u_1=u_2=u_3=u$, then a residual $Z_3$ symmetry exists, and the eigenvalues 
of ${\cal M}_\nu$ are simply $u^2(a+2b)$, $u^2(a-b)$, and $u^2(a-b)$. 
However, since the first eigenvalue corresponds to the eigenstate $(\nu_e + 
\nu_\mu + \nu_\tau)/\sqrt 3$, this is not a realistic scenario. Consider 
now the case
\begin{equation}
u_2 = u_3 = u \neq u_1.
\end{equation}
This makes ${\cal M}_\nu$ of the form advocated in Ref.~\cite{m02} and 
results in $\theta_{13} = 0$ and $\theta_{23}=\pi/4$.  Since $\theta_{13}=0$ 
implies that $CP$ is conserved in neutrino oscillations, the condition 
$u_2=u_3$ should be considered ``natural'' in the sense that it is 
protected by a symmetry.  Note that this alone does not imply 
$\theta_{23}=\pi/4$, which needs also $A_4$ for it to be true. 
[It certainly does not come from $\nu_\mu - \nu_\tau$ exchange as often 
suggested, because that would imply $\mu - \tau$ exchange as well, which 
cannot be sustained in the complete Lagrangian of the theory as a symmetry 
because $m_\mu \neq m_\tau$.]

Using the condition of Eq.~(7), ${\cal M}_\nu$ of Eq.~(5) can be rewritten 
as
\begin{equation}
{\cal M}_\nu = \pmatrix{\lambda^2 a & \lambda b & \lambda b \cr \lambda b & 
a & b \cr \lambda b & b & a}.
\end{equation}
In the basis $\nu_e$,  $(\nu_\mu+\nu_\tau)/\sqrt 2$, and $(-\nu_\mu+\nu_\tau)
/\sqrt 2$, this becomes
\begin{equation}
{\cal M}_\nu = \pmatrix{\lambda^2 a & \sqrt 2 \lambda b & 0 \cr \sqrt 2 
\lambda b & a+b & 0 \cr 0 & 0 & a-b},
\end{equation}
yielding one exact eigenvalue and eigenstate:
\begin{equation}
m_3 = a-b, ~~~ \nu_3 = (-\nu_\mu+\nu_\tau)/\sqrt 2.
\end{equation}
In the submatrix spanning $\nu_e$ and $(\nu_\mu+\nu_\tau)/\sqrt 2$, consider 
\begin{equation}
{\cal M}_\nu {\cal M}_\nu^\dagger = \pmatrix{|\lambda|^4|a|^2 + 
2|\lambda|^2|b|^2 & \sqrt 2 \lambda (|b|^2 + a^*b + |\lambda|^2 ab^*) \cr 
\sqrt 2 \lambda^* (|b|^2 + ab^* + |\lambda|^2 a^*b) & |a+b|^2 + 
2|\lambda|^2|b|^2}.
\end{equation}
The limit $|m_1|^2=|m_2|^2$ is reached if
\begin{equation}
|a+b|^2-|\lambda|^4|a|^2=0, ~~~ |b|^2+a^*b+|\lambda|^2ab^*=0,
\end{equation}
both of which are satisfied if $b=-a(1+|\lambda|^2)$.  In this limit, 
$\Delta m^2_{sol}=0$ and
\begin{equation}
\Delta m^2_{atm} \equiv |m_3|^2 - (|m_1|^2+|m_2|^2)/2 = 2 |a|^2 
(1-|\lambda|^4)(2+|\lambda|^2).
\end{equation}
To obtain a nonzero $\Delta m^2_{sol}$ and the value of $\theta_{12}$, 
consider
\begin{equation}
b = -a(1+|\lambda|^2+\epsilon),
\end{equation}
then
\begin{eqnarray}
\Delta m^2_{sol} &\equiv& |m_2|^2 - |m_1|^2 \nonumber \\ 
&=& |a|^2 [|(\epsilon + \epsilon^*)|\lambda|^2 + |\epsilon|^2|^2 + 
8 |\lambda|^2 |\epsilon^* + \epsilon |\lambda|^2 + |\epsilon|^2|^2]^{1/2},
\end{eqnarray}
and
\begin{equation}
\tan^2 2 \theta_{12} = {8 |\lambda|^2 |\epsilon^* + \epsilon |\lambda|^2 + 
|\epsilon|^2|^2 \over |(\epsilon + \epsilon^*)|\lambda|^2 + |\epsilon|^2|^2}.
\end{equation}

There are two natural limits of the parameter $\lambda$. (A) $\lambda=1$ 
corresponds to $u_1=u_2=u_3=u$, which is protected by a residual $Z_3$ 
symmetry as discussed already. (B) $\lambda=0$ corresponds to $m_{\nu_e}=0$ 
and the decoupling of $\nu_e$ from $\nu_\mu$ and $\nu_\tau$, which is 
protected by a chiral U(1) symmetry. Hence values of $\lambda$ near 1 and 0 
will be considered from now on.

(A) For $|\lambda| \simeq 1$, $\epsilon$ is expected to be small compared 
to it in Eq.~(14). In that case,
\begin{eqnarray}
\Delta m^2_{sol} \simeq 2 |a|^2 |\lambda| [(Re \epsilon)^2 (2+|\lambda|^2)
(1+2|\lambda|^2) + 2(Im \epsilon)^2 (1-|\lambda|^2)^2]^{1/2},
\end{eqnarray}
and
\begin{equation}
\tan^2 2 \theta_{12} \simeq 8 \left[ \left( {1+|\lambda|^2 \over 2|\lambda|} 
\right)^2 + {(Im \epsilon)^2 \over (Re \epsilon)^2} \left( {1-|\lambda|^2 
\over 2|\lambda|} \right)^2 \right].
\end{equation}
This means that $|\tan 2 \theta_{12}| > 2 \sqrt 2$, or equivalently 
$\sin^2 \theta_{12} > 1/3$, to be compared with the current experimental 
fit of $\sin^2 \theta_{12} = 0.31 \pm 0.03$.

Using the typical experimental values
\begin{equation}
\Delta m^2_{atm} = 2.5 \times 10^{-3}~{\rm eV}^2, ~~~ 
\Delta m^2_{sol} = 8.0 \times 10^{-5}~{\rm eV}^2,
\end{equation}
and assuming $\epsilon$ to be real, its value and those of $\sin^2 
\theta_{12}$ and $|\lambda^2 a|$ are given in Table 1.
\begin{table}[htb]
\caption{Values of $\sin^2 \theta_{12}$, $\epsilon$, and $|\lambda^2 a|$ as 
functions of $|\lambda|$.}
\begin{center}
\begin{tabular}{|c|c|c|c|}
\hline 
$|\lambda|$ & $\sin^2 \theta_{12}$ & $\epsilon$ & $|\lambda^2 a|$ \\ 
\hline
0.7 & 0.342 & 0.027 & 0.013~{\rm eV} \\ 
0.8 & 0.337 & 0.020 & 0.018~{\rm eV} \\ 
0.9 & 0.334 & 0.011 & 0.029~{\rm eV} \\ 
1.0 & 0.333 & -- & -- \\ 
1.1 & 0.334 & 0.014 & 0.035~{\rm eV} \\ 
1.2 & 0.336 & 0.032 & 0.026~{\rm eV} \\ 
1.3 & 0.338 & 0.055 & 0.023~{\rm eV} \\ 
1.4 & 0.341 & 0.082 & 0.021~{\rm eV} \\ 
\hline
\end{tabular}
\end{center}
\end{table}
It shows that $\sin^2 \theta_{12}$ is very near 1/3 and cannot be 
distinguished in practice from being exactly 1/3 \cite{hps}, as in some 
models. The last column corresponds to the expected value of the effective 
neutrino mass measured in neutrinoless double beta decay.

(B) For $|\lambda| \simeq 0$, consider $|\epsilon|$ also to be of order 
$|\lambda|$, then
\begin{eqnarray}
\Delta m^2_{atm} &\simeq& 4 |a|^2, \\
\Delta m^2_{sol} &\simeq& |a|^2 |\epsilon| \sqrt {|\epsilon|^2 + 
8 |\lambda|^2}, \\
\tan^2 2 \theta_{12} &\simeq& 8 |\lambda|^2/|\epsilon|^2.
\end{eqnarray}
In this case, $|a| = 0.025$ eV, and $\sin^2 \theta_{13} < 1/3$ can be 
obtained for $|\lambda| < |\epsilon|$.  Suppose it is fixed at 0.31, then 
$|\lambda| = 0.19$, $|\epsilon| = 0.22$, and $|\lambda^2 a| = 9.0 \times 
10^{-4}$ eV.

In conclusion, it has been shown in this paper that a new application of the 
non-Abelian discrete symmetry $A_4$ in the context of the canonical seesaw 
mechanism is successful in obtaining a realistic neutrino mixing matrix 
with $\theta_{13}=0$, $\theta_{23}=\pi/4$, and a prediction of $\sin^2 
\theta_{12}$ very near 1/3 in a particular symmetry limit. As Eq.~(13) 
shows, the normal (inverted) hierarchy of neutrino masses is obtained for 
$|\lambda|$ less (greater) than 1.  Typical values of the effective neutrino 
mass measured in neutrinoless double beta decay are given in Table 1.

I thank Steve King for useful discussions. This work was supported in part 
by the U.~S.~Department of Energy under Grant No. DE-FG03-94ER40837.

\bibliographystyle{unsrt}

\begin{thebibliography}{99}
\bibitem{mr01} E. Ma and G. Rajasekaran, Phys. Rev. {\bf D64}, 113012 (2001); 
E. Ma, Mod. Phys. Lett. {\bf A17}, 289 (2002); ibid. {\bf A17}, 627 (2002).
\bibitem{bmv03} K. S. Babu, E. Ma, and J. W. F. Valle, Phys. Lett. {\bf B552}, 
207 (2003); E. Ma, Mod. Phys. Lett. {\bf A17}, 2361 (2002); M. Hirsch, J. C. 
Romao, S. Skadhauge, J. W. F. Valle, and A. Villanova del Moral, Phys. Rev. 
{\bf D69}, 093006 (2004).
\bibitem{m0203} E. Ma, hep-ph/0208077; hep-ph/0208097; hep-ph/0307016; 
hep-ph/0311215.
\bibitem{m04} E. Ma, Phys. Rev. {\bf D70}, 031901(R) (2004); New J. Phys. 
{\bf 6}, 104 (2004); hep-ph/0409075.
\bibitem{m05} E. Ma, Mod. Phys. Lett. {\bf A20}, 1953 (2005).
\bibitem{af05} G. Altarelli and F. Feruglio, hep-ph/0504165; E. Ma, Phys. 
Rev. {\bf D72}, 037301 (2005).
\bibitem{cfm05} S.-L. Chen, M. Frigerio, and E. Ma, hep-ph/0504181, M. Hirsch, 
A. Villanova del Moral, J. W. F. Valle, and E. Ma, hep-ph/0507148.
\bibitem{bh05} K. S. Babu and X.-G. He, hep-ph/0507217.
\bibitem{seesaw} M. Gell-Mann, P. Ramond, and R. Slansky, in 
{\em Supergravity}, edited by P. van Nieuwenhuizen and D. Z. Freedman 
(North-Holland, Amsterdam, 1979), p.~315; T. Yanagida, in {\em Proceedings 
of the Workshop on the Unified Theory and the Baryon Number in the Universe}, 
edited by O. Sawada and A. Sugamoto (KEK Report No.~79-18, Tsukuba, Japan, 
1979), p.~95; R. N. Mohapatra and G. Senjanovic, Phys. Rev. Lett. {\bf 44}, 
912 (1980).
\bibitem{m02} E. Ma, Phys. Rev. {\bf D66}, 117301 (2002).
\bibitem{hps} P. F. Harrison, D. H. Perkins, and W. G. Scott, Phys. Lett. 
{\bf B530}, 167 (2002).
\end{thebibliography}

\end{document}